\documentclass[twocolumn,times]{aastex7}
\usepackage{hyperref}
\usepackage{natbib}
\usepackage{orcidlink}

\newcommand{\msun}{$M_{\odot}$}
\newcommand{\rsun}{$R_{\odot}$}

\newcommand{\kms}{km~s$^{-1}$}

\shorttitle{Photometric Morphology Changes in GD\,492}
\shortauthors{Dsouza et al.}

\graphicspath{{./}{figures/}}
\begin{document}

\title{Evidence for an Evolving Surface on the Partly Burnt Runaway Stellar Remnant GD\,492}

\author[0009-0006-2584-3292]{Reuben~Dsouza}
\affiliation{Department of Astronomy \& Institute for Astrophysical Research, Boston University, 725 Commonwealth Ave., Boston, MA 02215, USA}
\email{reubendz@bu.edu}

\author[0000-0001-5941-2286]{J.~J.~Hermes}
\affiliation{Department of Astronomy \& Institute for Astrophysical Research, Boston University, 725 Commonwealth Ave., Boston, MA 02215, USA}
\email[show]{jjhermes@bu.edu}

\author[0000-0002-2761-3005]{Boris~T.~G{\"a}nsicke}
\affiliation{Department of Physics, University of Warwick, Coventry CV4 7AL, UK}
\email{boris.gaensicke@gmail.com}

\author[0000-0002-9632-6106]{Ken~J.~Shen}
\affiliation{Department of Astronomy and Theoretical Astrophysics Center, University of California, Berkeley, CA 94720, USA}
\email{kenshen@astro.berkeley.edu}

\author[0009-0000-8806-5623]{Ciarra~Coston}
\affiliation{Department of Astronomy \& Institute for Astrophysical Research, Boston University, 725 Commonwealth Ave., Boston, MA 02215, USA}
\email{ccoston@bu.edu}

\author[0000-0002-9090-9191]{Roberto~Raddi}
\affiliation{Departament de F\'{i}sica, Universitat Polit\`{e}cnica de Catalunya, c/Esteve Terrades 5, 08860 Castelldefels, Spain}
\email{roberto.raddi@upc.edu}

\author[0000-0003-0089-2080]{Mark~Hollands}
\affiliation{Department of Physics, University of Warwick, Coventry CV4 7AL, UK}
\email{mark.hollands@warwick.ac.uk}

\begin{abstract}

GD\,492 (LP 40-365) is a partly burnt, hypervelocity runaway stellar remnant thought to have survived a thermonuclear supernova. We analyze 12 sectors of TESS photometry spanning more than four years to investigate the previously reported 8.9-hr photometric modulation of the star. We find that the modulation changes with time, with individual-sector fits yielding apparent periods ranging from 8.8606 to 8.9048 hr (0.5\% changes), more than 25 times the formal measurement uncertainties. Both the period and shape of the modulation evolve on a timescales of approximately one to two months. We show qualitatively that a two-spot model with evolving surface structure can reproduce several of the observed light-curve morphologies, although the geometry of the spots is not uniquely constrained. These observations suggest that changing surface inhomogeneities are a plausible origin of the photometric variability. Differential rotation and magnetic activity may contribute to this evolution, potentially providing new constraints on the outer convective structure of GD\,492 and on the interpretation of its atmospheric abundances.

\end{abstract}

\section{Introduction} \label{sec:intro}

GD\,492 (LP 40-365) is the prototype of a class of partly burnt, hypervelocity stellar remnants thought to have survived thermonuclear supernova explosions \citep{GD492Discovery}. It is currently escaping the Galaxy at a velocity of roughly 850\,\kms\ and has an unusual atmosphere dominated by oxygen and neon, with no detectable hydrogen or helium \citep{RaddiGD492,FurtherInsight}.

Its low mass of $\sim$\,0.28\,\msun, high velocity, and peculiar atmosphere \citep{RaddiNewGD492} suggest that GD\,492 is the bound remnant of a white dwarf that was only partially disrupted during a thermonuclear explosion, potentially related to the Type Iax class of supernovae \citep{FurtherInsight}. A leading hypothesis is that it formed in a binary system when a near-Chandrasekhar-mass white dwarf underwent an asymmetric deflagration that ejected a high-velocity remnant, named as LP 40-365 objects \citep{Shen2025D6}. At least eight suspected LP 40-365 type runaway stars are now known \citep{GD492Discovery,El-Badry2023,RaddiNewGD492,Bhat2026,El-Badry2026}.

Using four sectors (Sectors 14, 15, 20, and 21) of photometry from the Transiting Exoplanet Survey Satellite (TESS), \citet{HermesGD492} reports a periodic modulation in GD\,492 at 8.914\,hr with an amplitude of 1.0\%. A commensurate 8.9-hr signal at 5.8\% is also observed in Hubble Space Telescope ultraviolet data. The chromatic amplitude ratio suggests the modulation is caused by a surface spot rotating in and out of view, similar to other spots on magnetic white dwarfs \citep{MagWDVar,Nonsin1}.

\begin{figure*}[t!]
    \centering
    \includegraphics[width=\linewidth]{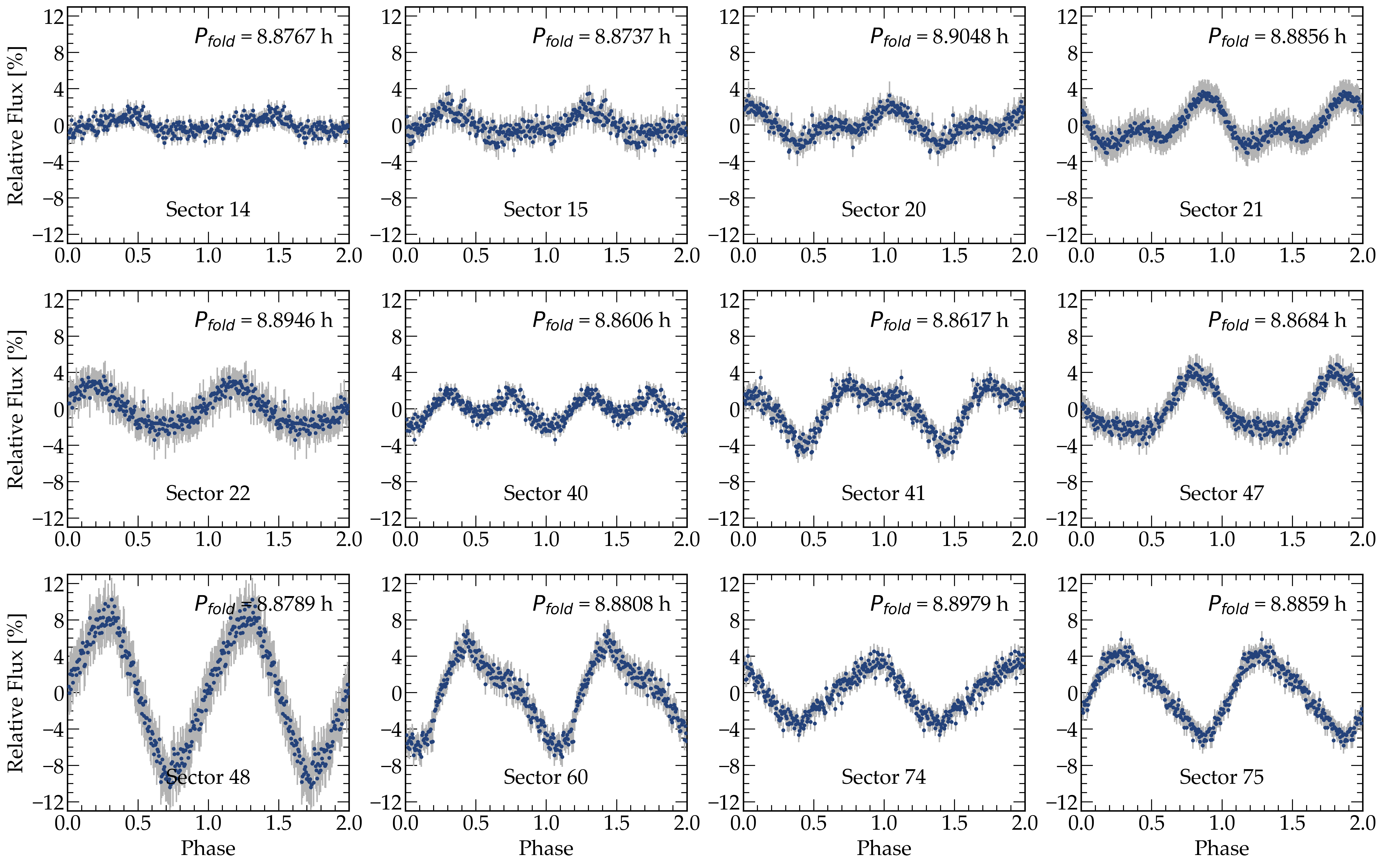}
    \caption{Phase-folded TESS light curves of GD\,492 for all 12 observing sectors collected as of 2024~March. The amplitude of the 8.9-hr photometric modulation varies by up to a factor of 10 between epochs, with the waveform varying between adjacent sectors like Sector 47 and 48. The folding period (at top right) is the best-fit period for that sector (see Table~\ref{tab:sectors}) and differs for each subplot.}
    \label{fig:all_sec}
\end{figure*}

The light curve does not appear purely sinusoidal and shows another variability near the second harmonic of the fundamental frequency. The comparatively slow 8.9-hr modulation is interpreted as evidence that GD\,492 is the partially disrupted remnant rather than the surviving donor of the progenitor binary system; its radius has expanded by at least a factor of 40, and if it conserved angular momentum, it would have had a rotation period, P $<$\,2\,min when it was a massive ONe-core white dwarf \citep{HermesGD492}.

The initial TESS observations cover less than an 8-month baseline. With more monitoring of GD\,492, TESS reveals that the dominant modulation varies in both best-fitting period and morphology. Here we analyze 12 sectors of TESS photometry of GD\,492 spanning more than four years. We quantify the sector-to-sector evolution of the 8.9-hr modulation in period, amplitude, and light-curve morphology, and examine its evolution within adjacent TESS sectors. We then use a qualitative two-spot model to test whether evolving surface structure can reproduce several of the observed light-curve shapes. Finally, we consider possible physical origins of this evolution, including differential rotation, pulsations, magnetic activity, and the possible implications for the outer structure of GD\,492.

\section{TESS Full-Frame Image Observations} \label{sec:tess}

GD\,492 (TIC 198510602, $T$\,=\,15.5\,mag) has been observed by TESS \citep{Ricker2015} in 12 sectors between 2019~July and 2024~February. Each TESS sector provides up to 27.5 days of near-continuous monitoring.

Data from Sectors 14, 15, 20, 21, and 22 was obtained from 30-minute full-frame images (FFIs). Sectors 40, 41, 47, and 48 has been obtained using 10-minute FFIs, and Sectors 60, 74, and 75 use 200-second FFIs. In total, the 12 sectors contain $44{,}580$ photometric measurements spanning more than four years.All the TESS data used in this paper can be found in MAST: \dataset[https://doi.org/10.17909/gdv3-fh23]{https://doi.org/10.17909/gdv3-fh23}.

We extract light curves for all 12 sectors using the TESS-Gaia Light Curve (\verb|TGLC|) package \citep{BrandtTGLC}. \verb|TGLC| uses Gaia DR3 source positions and fluxes as priors in point-spread-function photometry of the TESS full-frame images, helping to reduce contamination from nearby sources.

\begin{table*}[]
\centering
\begin{tabular*}{\textwidth}{c @{\extracolsep{\fill}} c ccccccc}
\hline \hline
Sector & Data collection period & Cadence (s) & $N$ & $P_1$ (hr) & $A_1$ (\%) & $A_2$/$A_1$ & $\Delta P_2$ (s) & $A_{2,\mathrm{fit}}$ (\%) \\ \hline
 14 & 2019 Jul 18 -- 2019 Aug 14 & 1800 & 1237 & 8.8767(33) & 0.770(51) & 0.423 & $-$20(11) & 0.345(55)  \\
 15 & 2019 Aug 15 -- 2019 Sep 10 & 1800 & 1190 & 8.8737(22) & 1.248(56) & 0.506 & 20.0(7.2) & 0.673(67) \\
 20 & 2019 Dec 24 -- 2020 Jan 20 & 1800 & 1065 & 8.9048(14) & 1.452(53) & 0.689 & 0.7(4.7) & 1.010(70) \\
 21 & 2020 Jan 21 -- 2020 Feb 18 & 1800 & 1037 & 8.88558(68) & 2.935(58) & 0.773 & $-$9.9(2.9) & 2.29(11) \\
 22 & 2020 Feb 19 -- 2020 Mar 17 & 1800 & 808 & 8.8946(16) & 2.388(62) & 0.187 & $-$110.0(9.5) & 0.68(10) \\ \hline
 40 & 2021 Jun 24 -- 2021 Jul 23 & 600 & 3917 & 8.8606(10) & 0.779(47) & 1.796 & 6.5(2.7) & 1.411(48) \\
 41 & 2021 Jul 23 -- 2021 Aug 20 & 600 & 3658 & 8.8617(12) & 2.429(62) & 0.499 & $-$5.0(4.2) & 1.217(74) \\
 47 & 2021 Dec 30 -- 2022 Jan 28 & 600 & 2998 & 8.86838(87) & 3.577(55) & 0.309 & 36.0(4.3) & 1.289(85) \\
 48 & 2022 Jan 28 -- 2022 Feb 26 & 600 & 2465 & 8.87891(56) & 7.392(63) & 0.107 & 20(12) & 0.84(16) \\ \hline
 60 & 2022 Dec 23 -- 2023 Jan 18 & 200 & 6938 & 8.88083(75) & 4.520(61) & 0.386 & 11.0(3.1) & 1.904(82) \\
 74 & 2024 Jan 03 -- 2024 Jan 30 & 200 & 8860 & 8.8979(12) & 3.052(55) & 0.142 & $-$44.0(7.2) & 0.566(64) \\
 75 & 2024 Jan 30 -- 2024 Feb 26 & 200 & 10407 & 8.88585(76) & 4.352(55) & 0.193 & $-$53.9(3.1) & 1.419(69) \\
\hline \hline
\end{tabular*}
    \caption{Individual TESS-sector light curve analysis. Values in parentheses reflect the formal uncertainties returned by the nonlinear least-squares fit. $N$ is the number of photometric measurements in each sector. $A_1$ is the amplitude of the fitted fundamental at $f_1$, and $A_2$ is the amplitude of the component fixed at $2f_1$. $A_{2,\mathrm{fit}}$ is the amplitude of independently fitted period of the shorter-period signal, with the offset $\Delta P_2 = P_{2,\mathrm{fit}}-P_1/2$.}
    \label{tab:sectors}
\end{table*}

This is particularly useful for GD\,492 because of its relatively faint TESS magnitude and because the earlier work of \citet{HermesGD492} using Sectors 14, 15, 20, and 21 analyzed \verb|eleanor| light curves that did not explicitly correct for crowding \citep{Eleanor1}, with image cutouts created using \verb|Astrocut| \citep{Eleanor2}. We therefore re-extract the earlier sectors together with all new observations using a uniform photometric pipeline.

We retain only observations with both the TESS and \verb|TGLC| quality flags equal to zero and use the calibrated PSF photometry provided by \verb|TGLC|. Each light curve is normalized independently to the median flux of each sector. We assign all observations within a given sector the same sector-specific \verb|TGLC| PSF-flux uncertainty. We do not apply any additional outlier rejection or long-term detrending.

\section{Evolution of the TESS Photometry} \label{sec:evolution}

Figure~\ref{fig:all_sec} shows the phase-folded light curves of GD\,492 for all 12 TESS sectors. The 8.9-hr modulation is clearly not stable across the observing baseline: the amplitude varies by up to a factor of 10, and the waveform varies in shape between sectors. 

\begin{figure}[t!]
    \centering
    \includegraphics[width=\linewidth]{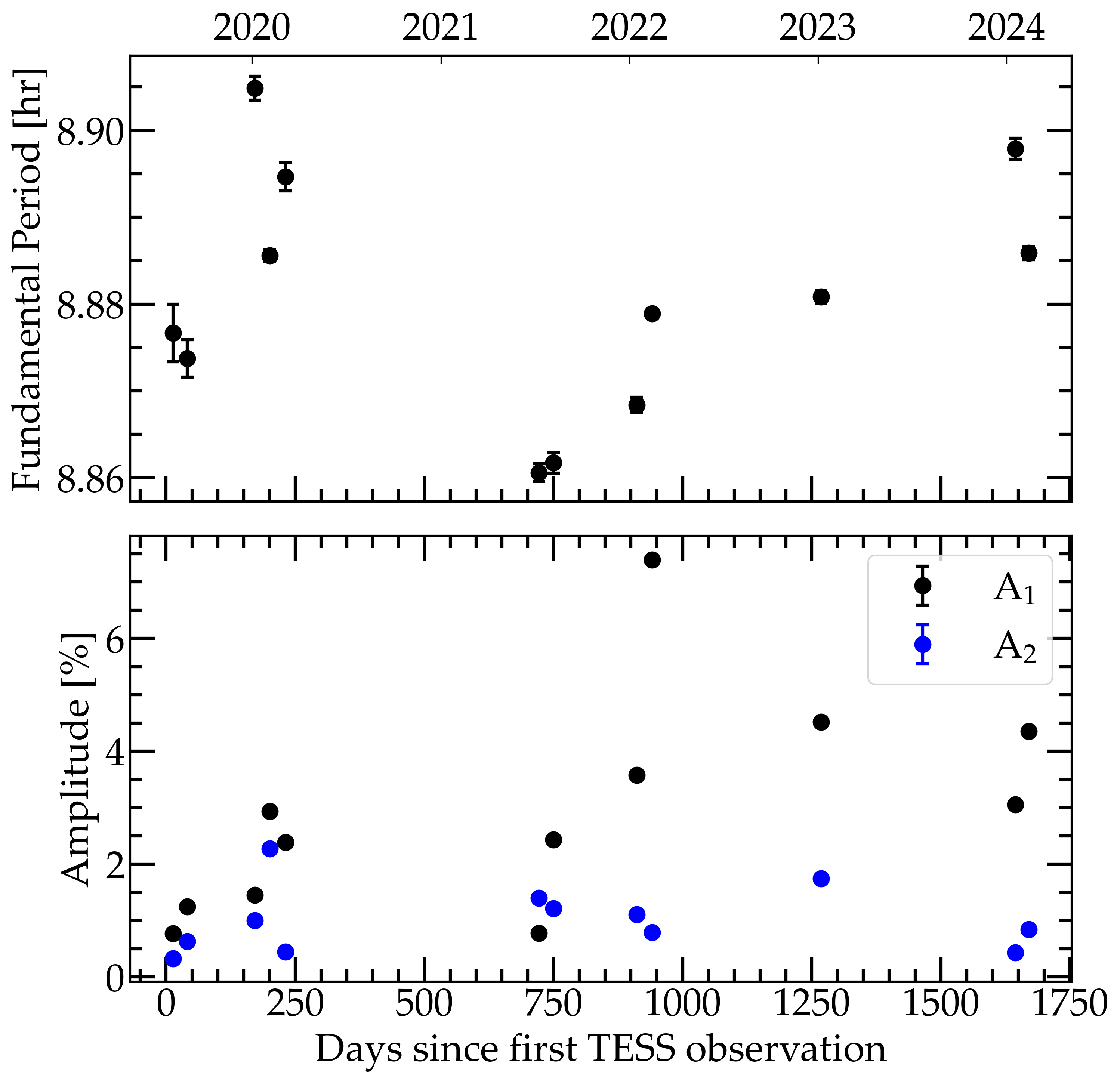}
    \caption{Best-fitting photometric periods (top panel) and amplitudes (bottom panel) measured independently in each TESS sector for GD\,492. Both quantities vary non-monotonically over the multi-year observing baseline. Error bars show the formal uncertainties returned by the sinusoidal fits and do not account for evolution of the light-curve waveform.}
    \label{fig:periods}
\end{figure}

\subsection{Sector-to-Sector Period \& Amplitude Variability}
\label{sec:sectorchanges}

For each TESS sector we use a Lomb-Scargle periodogram \citep{LSP} to identify the dominant frequency near the 8.9-hr modulation. We then refine the frequency, amplitudes, and phases using a nonlinear least-squares fit of the form

\begin{equation}
    y(t)=A_1\sin\left[2\pi\left(f_1t+\phi_1\right)\right]
+A_2\sin\left[2\pi\left(2f_1t+\phi_2\right)\right],
\end{equation}

where $f_1$ is the fitted fundamental frequency and the second component is fixed at exactly $2f_1$. The initial guesses for the frequency and amplitudes come from the periodogram; we initialize both phases at $\phi_1=\phi_2=0.5$.

\begin{figure*}[t!]
    \centering
    \includegraphics[width=\linewidth]{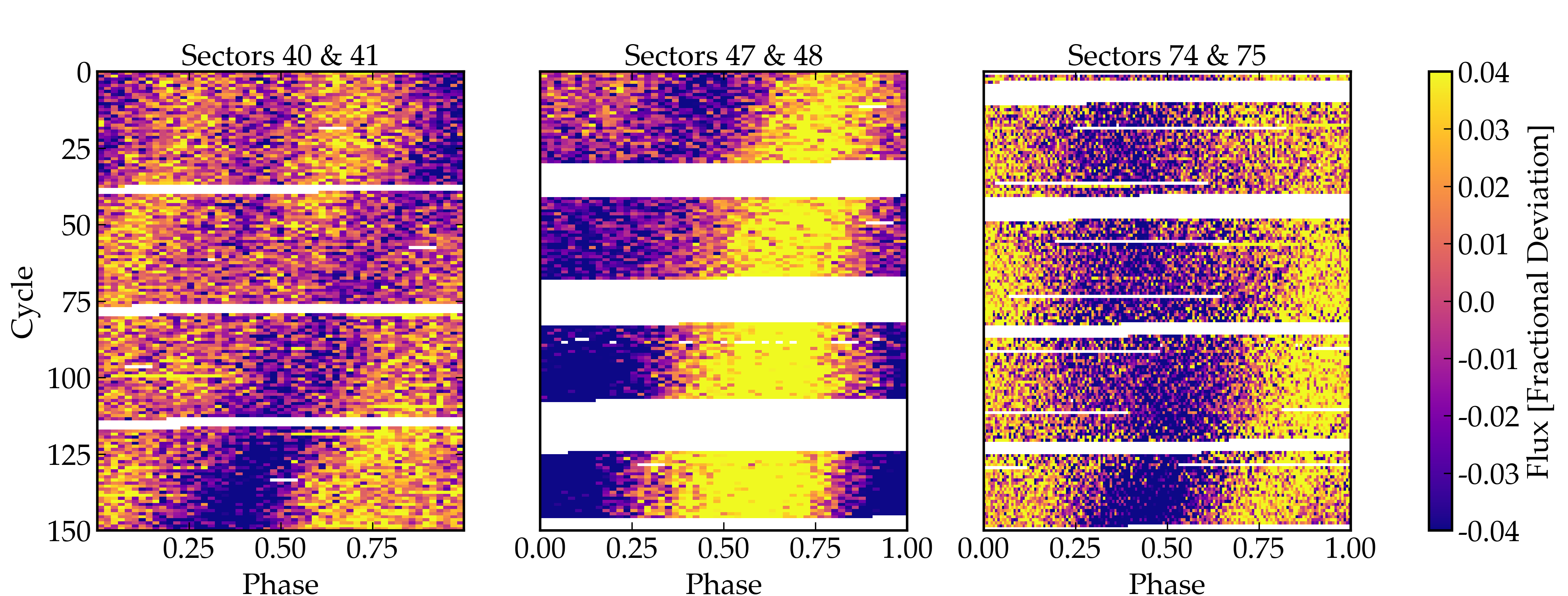}
    \caption{River plots (trailed light curves) in which successive photometric cycles are displayed as a function of phase and time, for adjacent TESS Sectors 40--41, 47--48, and 74--75, phased at the global best-fitting photometric period of 8.892803\,hr. The amplitude and morphology of the 8.9-hr modulation changes within these adjacent sectors, demonstrating evolution on month-long timescales.}
    \label{fig:river}
\end{figure*}

The earlier analysis of GD\,492 based on the first four sectors of observations reported a period of $8.91405(35)$\,hr and an amplitude in TESS of $1.038(42)$\% \citep{HermesGD492}. Using 12 sectors, the global best-fitting photometric period is $8.892803(11)$\,hr. We perform a Monte Carlo uncertainty calculation by perturbing the observed flux measurements according to their photometric uncertainties and repeating the period and amplitude fits. The resulting uncertainties are approximately 10\% larger than the formal least-squares uncertainties. This comparison suggests that the formal uncertainties adequately characterize the statistical precision of the fits under the adopted noise model. The uncertainties quoted throughout the manuscript are the formal least-squares uncertainties, but neither method fully accounts for systematic effects associated with the evolving light-curve morphology.


The resulting sector-by-sector measurements are listed in Table~\ref{tab:sectors} and shown in Figure~\ref{fig:periods}. The fitted photometric period ranges from $8.8606(10)$\,hr in Sector 40 to $8.9048(14)$\,hr in Sector 20, a total range of 0.0442\,hr or approximately 0.50\% of the fundamental period, more than 25 times larger than the formal measurement uncertainties. The evolution does not exhibit a monotonic increase or decrease with time, arguing against a simple model of a constant secular period derivative.

The fitted fundamental amplitude varies even more dramatically in comparison to the measured period, from less than 0.8\% in Sector 14 to nearly 7.4\% in Sector 48. The harmonic signal also varies strongly, which we report as the ratio of the harmonic amplitude to the fundamental amplitude ($A_2/A_1$) in Table~\ref{tab:sectors}. In Sector 40, the harmonic component fixed at $2f_1$ is actually considerably stronger than the fundamental.

Our uniform \verb|TGLC| extraction shows that even during the early TESS observations the best-fit period and amplitude differed between epochs. Additional sectors demonstrate that this variability persists throughout all four years of TESS observations.

\subsection{The Second Harmonic}
\label{sec:harmonic}

The full TESS data set of all 12 sectors shows substantial power near twice the dominant frequency, but this signal may not be well described as a strictly coherent harmonic of the global best-fitting period. Fitting the full data set with the second component constrained to exactly $2f_1$ gives a fundamental period of 8.892803\,hr and amplitudes of $A_1=1.557(20)$\% and $A_2=0.063(20)$\%. The exact half-period associated with this fit is $4.446402$ hr.

As in \citet{HermesGD492}, we interpret the second harmonic as a consequence of the non-sinusoidal waveform, which should generate a signal at integer harmonics of the fundamental. When instead fitting the short-period signal independently near $2f_1$, we obtain a period of $4.4267211(71)$\,hr with an amplitude of $0.642(21)$\%. The independently fitted signal is displaced from the exact harmonic by approximately $-$0.01968\,hr, over 100 times larger than the formal fitting uncertainties.

\begin{figure*}[t!]
    \centering
    \includegraphics[width=\linewidth]{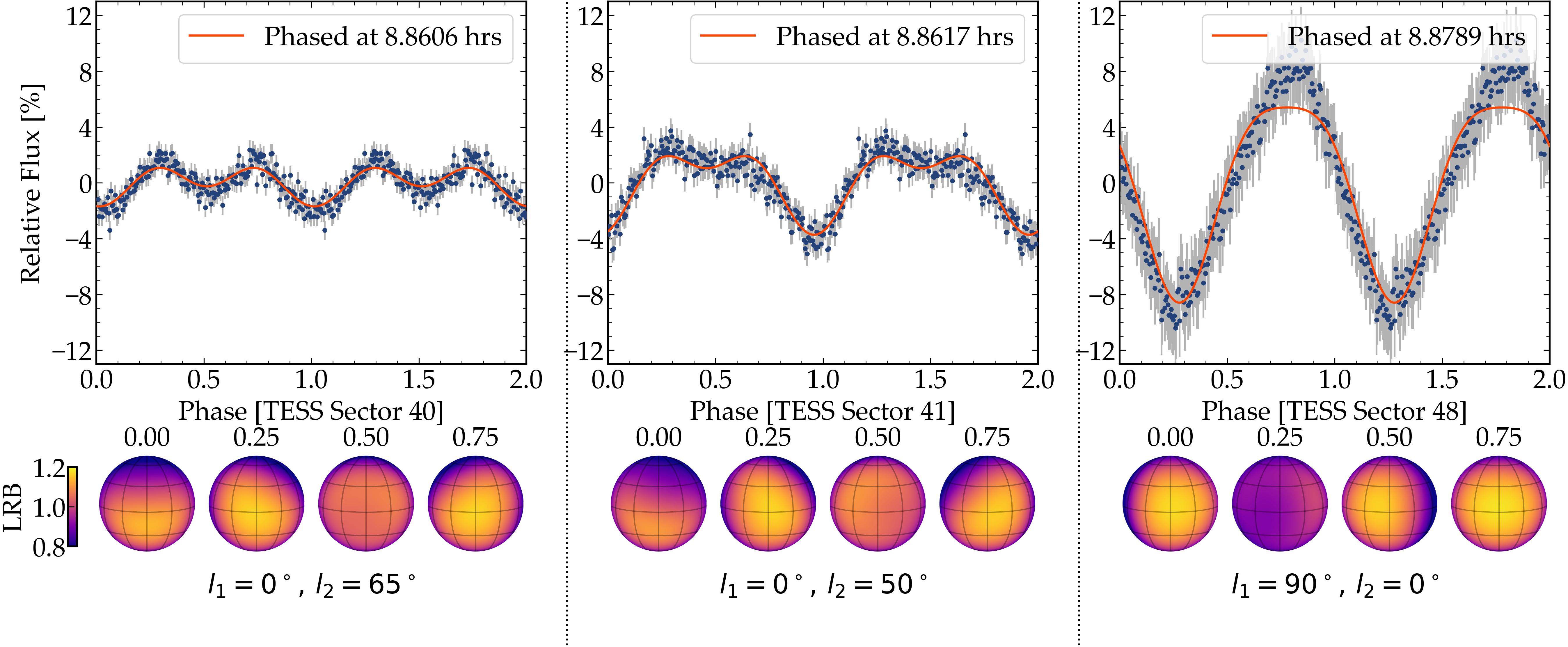}
    \caption{Phased light curves of GD\,492 for Sectors 40, 41, and 48, folded at 8.8606 hours, 8.8617 hours, and 8.8789 hours, respectively. The red curve is obtained from models of a representative star with two spots shown in the lower panel, where LRB refers to the local relative brightness for the model star. We are able to qualitatively reproduce the light curves for all three sectors using the same spot sizes, contrasts, and longitudinal separation, while only changing the latitude of the spots, $l_A$ and $l_B$.}
    \label{fig:model}
\end{figure*}

Since the light-curve morphology itself evolves, this offset does not necessarily imply the presence of two stable coherent frequencies. Instead, it motivates examining how the near-$2f_1$ structure changes from sector to sector. Table~\ref{tab:sectors} shows the offset measured when comparing the independently fitted signal from the exact harmonic. The offset ranges from $-$110.0(9.5) s in Sector 22 to +36.0(4.3) s in Sector 47, with most sectors having offsets larger than the formal fitting uncertainties. The changing sign and magnitude of these offsets are consistent with an evolving spot pattern, since spot growth and decay, migration, and differential rotation can modulate the rotational waveform and produce Fourier power near but not exactly at $2f_1$ \citep{Balaji2015}. 


\subsection{Evolution Within Adjacent TESS Sectors}
\label{sec:riverplot}

To examine changes on timescales shorter than the full multi-year baseline, we construct river plots for three epochs: the adjacent Sectors 40--41, 47--48, and 74--75 in Figure~\ref{fig:river}. These river plots are trailed light curves in which successive photometric cycles are displayed as a function of phase and time, allowing changes in the amplitude and waveform of the modulation to be followed within consecutive observing sectors (roughly two months). The cycles correspond to successive rotation cycles, with 0 being the beginning of the observations and 150 the end. All plots are phase-folded at the global best-fit period of 8.892803\,hr.

The three epochs show qualitatively different behavior. Across Sectors 40--41, the phase of maximum light shifts by roughly 0.7 cycles over approximately two months while the waveform also changes from having two bright spots to one. Across Sectors 47--48, the modulation increases in amplitude while its maximum shifts by roughly 0.3 cycles. In contrast, Sectors 74--75 show a changing amplitude but a comparatively stable phase. Thus, the modulation does not evolve at a single uniform rate: different epochs show different combinations of phase, amplitude, and waveform evolution on month-long timescales.

A light curve folded at a fixed reference period will show systematic phase drift whenever its local photometric period differs from the adopted folding period, so we do not interpret the observed phase shifts as evidence for periodic evolution of the photometric modulation. Instead, the river plots show that the amplitude and light-curve morphology change within adjacent sectors, demonstrating that the photometric modulation evolves on month-long timescales and does so differently at different epochs.

\section{Qualitative Surface-Spot Modeling}
\label{sec:spotmodeling}

To test whether evolving surface inhomogeneities can plausibly reproduce the changing light-curve morphology of GD\,492, we construct simple two-spot models using \verb|starry| \citep{starry}, currently implemented in the \verb|jaxoplanet| framework \citep{jaxoplanet}. Our goal is not to infer a unique surface map, but rather to determine whether relatively simple changes in surface-brightness inhomogeneities can qualitatively reproduce several representative TESS light curves. We focus on Sectors 40, 41, and 48 because they span different observed morphologies, including the unusually strong near-half-period structure in Sector 40 and the more sinusoidal modulation in Sector 48.

The synthetic light curves require assumptions about stellar inclination, limb darkening, and spot properties. To obtain limb darkening coefficients, we compute a model atmosphere for GD\,492 using the code of \citet{Detlev} adopting the parameters of $T_{\rm eff}$\,=\,9800\,K and $\log g$\,=\,5.5 \citep{RaddiNewGD492}. The spectra are computed at nine discrete values of $\mu$, defined as $\cos(\gamma)$, where $\gamma$ is the angle between our line of sight and the normal of the surface at any point. We then calculate the TESS fluxes, I($\mu$), for each angle by integrating over the TESS bandpass. Finally, we fit the quadratic limb-darkening law to I($\mu$)/I(1) to get the two coefficients, $(a,b)=(0.1568,0.3328)$.

We find we can generate qualitative matches to several observed waveforms to multiple sectors if we adopt an inclination of $i=80^\circ$ with respect to the rotation axis of GD\,492, while noting that the inferred spot geometry is strongly degenerate with this choice of stellar inclination.

We use two spots (Spot A and Spot B) separated by a fixed $180^\circ$ in longitude, with angular radii of 0.7\,radians and 0.9\,radians and contrasts of 0.5 and 0.9 for Spots A and B, respectively. We define the contrast as $c = 1 - b_{ii}/b_{i}$, where $b_{ii}$ is the brightness at the spot center and $b_{i}$ is the brightness of the unspotted stellar surface, such that increasing $c$ corresponds to a darker spot \citep{contrast}. We hold these properties fixed between the three example sectors, and allow only the spot latitudes to vary.

Figure~\ref{fig:model} shows representative configurations that qualitatively reproduce the observed light curves. For Sector 40, an illustrative configuration places Spot A at $0^\circ$ latitude and Spot B at $65^\circ$ latitude. For Sector 41, Spot A remains at $0^\circ$ but Spot B is at $50^\circ$ latitude. For Sector 48, Spot A is at $90^\circ$ latitude and Spot B at $0^\circ$ latitude.

These configurations demonstrate that changes in surface geometry alone (specifically spot latitude) can produce substantially different light curve shapes and modulation while holding the adopted spot sizes and contrasts fixed. However, the solutions are not unique, and similar models do not reproduce every TESS sector without allowing additional parameters such as spot size, contrast, longitude, or number of surface features to evolve. These models therefore demonstrate that evolving surface inhomogeneities can plausibly reproduce the observed morphology changes, but require more degrees of freedom for more inferences to be drawn from them. We connect these toy models to a possible interpretation in the next section.

\section{Discussion} \label{sec:discussion}

\subsection{Origin of the Evolving Photometric Modulation}
\label{sec:origin}

The sector-to-sector period changes are unlikely to represent variations in the bulk rotation rate of GD\,492. The full range of period change corresponds to $\Delta P/P$\,$\simeq$\,0.5\%. If this variation were produced by a global change in stellar radius while approximately conserving angular momentum, $P\propto R^2$ would require $\Delta R/R$\,$\simeq$\,0.25\%. For $R\simeq0.16$\,\rsun \citep{RaddiNewGD492}, this corresponds to a global radius change of roughly 280\,km. Such a structural change would require a change in the gravitational binding energy of order a few $\times10^{45}$\,erg for a $0.28$\,\msun\ star. The ratio of this energy to the present luminosity of 0.2\,$L_{\odot}$ corresponds to a characteristic energetic timescale greater than $100{,}000$\,yr, illustrating how implausible it is for the observed month-scale period variations to reflect global structural evolution.

Another possible explanation for the modulation observed in the light curves of GD\,492 is free precession. If GD\,492 possesses a persistent deformation (for example, one produced by an internal magnetic field), and if the rotation axis is misaligned with the principal axis of inertia, the star can undergo free precession \citep{Precession1,Precession2}. Free precession would periodically change the viewing geometry of surface spots. However, the TESS light curves do not show any obvious repeating periods of morphological changes. Without a quantitative prediction for the precession period and stellar deformation of GD\,492, we cannot exclude this mechanism, but the current observations provide no clear evidence for a coherent precession cycle.

Stellar oscillations are another possible explanation for photometric signals that evolve in period and amplitude, and pulsational driving cannot be ruled out solely from the unusual composition of GD\,492. However, the observed photometric timescales are difficult to reconcile with pulsations. Using $M\simeq0.28$\,\msun\ and $R\simeq0.16$\,\rsun, the dynamical timescale of GD\,492 is t $<$ 200\,s, orders of magnitude shorter than the observed 8.9-hr modulation. Models of hydrogen-poor stars can become unstable to radial pulsations through partial-ionization driving \citep{JefferySaio2016}, but the expected radial-mode periods are of order several hundred seconds rather than tens of thousands of seconds. An 8.9-hr $g$-mode would require a very high radial order, and we might also expect variations from other adjacent $g$-modes, since known extremely low-mass white dwarf $g$-mode pulsators are generally multiperiodic \citep{Hermes2013}. We therefore consider pulsations less likely than rotational modulation associated with evolving surface structure.

The most direct interpretation of the TESS observations is therefore that the surface pattern producing the 8.9-hr modulation itself evolves. The fundamental amplitude changes by nearly a factor of ten, the relative harmonic content changes strongly between sectors, and the phase-folded morphology can change substantially over month-long intervals. The qualitative models in Section~\ref{sec:spotmodeling} demonstrate that relatively simple changes in surface-brightness geometry can reproduce several of these morphologies.

\subsection{Differential Rotation and Magnetic Activity}
\label{sec:diffrot}

If different surface structures dominate the photometric modulation at different epochs, and those structures occur at different stellar latitudes, the range of measured periods could be explained by latitudinal differential rotation. Following \citet{Distefano2016}, we can estimate the magnitude of the corresponding photometric shear by defining
\begin{equation}
    \Delta\Omega_{\rm phot}
=
2\pi\left(\frac{1}{P_{\rm min}}-\frac{1}{P_{\rm max}}\right).
\end{equation}

Using $P_{\rm min}=8.8606$\,hr and $P_{\rm max}=8.9048$\,hr gives $\Delta\Omega_{\rm phot}\simeq0.084$\,rad\,d$^{-1}$, corresponding to a relative shear of $\Delta\Omega_{\rm phot} / \Omega~\simeq~0.005.$ The associated beat timescale \citep{Collier2002}, $2\pi/\Delta\Omega_{\rm phot}$, is approximately 74\,days. The corresponding lap timescale is comparable to the month-long evolution seen in some adjacent TESS sectors.

Relatively rapid spot morphology changes may indicate a magnetic activity timescale in GD\,492. Spectroscopy places an upper limit of approximately 20\,kG on the surface magnetic field from the absence of detectable Zeeman splitting \citep{HermesGD492}, but this limit does not exclude weaker fields capable of influencing the outer atmosphere. Rapidly rotating stars with convective envelopes commonly show evolving magnetic surface structure \citep{CoolSpots,DynamoLateStar}, although direct analogy with ordinary late-type stars should be used with caution given the unusual evolutionary history of GD\,492.

If the evolving surface structures are magnetic, an active dynamo provides one possible interpretation of the photometric behavior of the TESS observations. The toy models in Section~\ref{sec:spotmodeling} do not robustly measure the spot latitudes, so we do not attempt to use them to infer the magnitude or sign of the true equator-to-pole differential rotation from the current data; more detailed spot modeling could potentially constrain both.

\subsection{Implications for Convection in GD\,492}
\label{sec:convection}

If the evolving surface structure of GD\,492 is magnetic in origin, its rapid evolution may provide information about the poorly constrained outer structure of the star. One possible explanation for evolving magnetic surface structure is an active convection-driven dynamo \citep{DynamoLateStar}. However, the presence of magnetic spots would not uniquely establish this mechanism, and neither the existence nor the depth of a convective region in GD\,492 is currently observationally constrained.

Evolutionary calculations of partially disrupted remnants of thermonuclear supernovae predict structures that differ from ordinary white dwarfs and core-burning hot subdwarfs \citep{ModelConvection}. In particular, \citet{Shen2025D6} modeled the LP 40-365 stars as initially fully convective, low-mass oxygen-neon-rich remnants whose internal structure evolves as they contract through their Kelvin-Helmholtz cooling phase. In this scenario, the progenitor was an oxygen-neon white dwarf that approached the Chandrasekhar mass and underwent an oxygen deflagration that asymmetrically ejected most of the stellar mass and imparted a large kick to the surviving remnant.

The subsequent internal structure of such a remnant may evolve with time. Thus, even if GD\,492 evolved through an initially fully convective phase, it may not be fully convective at its present temperature and age. For example, analogous carbon-oxygen-rich models in \citet{Shen2025D6} eventually develop radiative cores as they contract toward the white dwarf cooling sequence. The present photometric variability therefore motivates more detailed calculations of convection and magnetic activity appropriate specifically to GD\,492.

Constraining the depth of any outer convection zone would have broader implications for how the atmospheric composition of GD\,492 is interpreted. The LP 40-365 stars usually have neon-dominated atmospheres enriched in intermediate-mass and iron-group elements \citep{GD492Discovery,RaddiNewGD492,Bhat2026,El-Badry2026}. In GD\,492 specifically, the measured super-solar manganese-to-iron abundance of $[$Mn/Fe$]$\,=\,0.82 dex has been used to argue for high-density burning in a near-Chandrasekhar-mass progenitor \citep{FurtherInsight}. However, translating these photospheric abundances into the bulk composition of the surviving remnant depends on how deeply the observable atmosphere is mixed with the stellar interior. Constraining the convection-zone depth in GD\,492 is therefore critical for determining whether its photospheric abundances can more directly connect to predicted Type Iax nucleosynthetic yields.

In the case of a fully convective remnant, this comparison becomes particularly constraining, since the atmospheric abundances should directly reflect the bulk composition. Existing ONe-deflagration models show discrepancies with the observed light- and intermediate-element abundances of GD\,492, although those calculations adopt a simplified initial oxygen-neon composition and omit several compositional layers and products of the prior evolution of the object \citep{Jones2019,RaddiNewGD492}.

\section{Conclusion} \label{sec:conclusion}

We analyze 12 sectors of TESS photometry of GD\,492 spanning more than four years and find that its previously reported 8.9-hr photometric modulation evolves significantly with time. The best-fitting photometric period shifts from as low as 8.8606\,hr to as high as 8.9048\,hr, a total range of approximately 0.5\% of the global-best-fit period, while the fundamental amplitude varies by nearly an order of magnitude. Adjacent TESS sectors show that the amplitude and morphology can evolve on month-long timescales. These changes are non-monotonic and therefore do not resemble a simple secular evolution of the photometric modulation.

A qualitative two-spot model demonstrates that evolving surface-brightness structure (especially changes in latitude) can reproduce several of the observed light-curve morphologies, although the surface geometry is not uniquely constrained. Evolving magnetic surface structure, potentially coupled with latitudinal differential rotation, could plausibly account for the observed photometric period changes.

The presence of evolving spot modulation from magnetic activity in GD\,492 would be a novel discovery. While magnetic cycles are common among active dwarfs \citep{Baliunas1995,Suarez2016}, several well-studied magnetic white dwarfs show highly coherent rotational modulation over months to years (e.g., \citealt{Brinkworth2005,Walters2021}).

Magnetic activity might imply an active dynamo, although no magnetic field is detected at the surface of GD\,492 to a limit of B $<$ 20\,kG due to a lack of observed Zeeman splitting \citep{HermesGD492}. Importantly, the absence of detectable Zeeman splitting does not preclude magnetically induced surface structure. Dynamo models of fully convective stars can generate magnetic activity with characteristic fields of only a few kG \citep{FCModel1,FCModel2}. More sensitive follow-up looking for magnetism in GD\,492 could potentially add constraints to the surface structure.

A convection-driven dynamo would likely require a large convection zone, with potentially important implications for the interpretation of the surface abundances of GD\,492. If convection mixes a large fraction of the stellar mass, and diffusion out of that mixed region occurs on a timescale much longer than the $\sim$\,5\,Myr flight time, the photospheric abundances could trace the composition of a large fraction of the remnant. In the limiting case of an effectively fully convective star, they could provide direct access to the abundance ratios of material processed during the near-Chandrasekhar-mass thermonuclear explosion.

\section*{Acknowledgments}

This material is based upon work supported by the National Aeronautics and Space Administration under Grant No. 80NSSC23K1068 issued through the Science Mission Directorate. This project has received funding from the European Research Council (ERC) under the European Union’s Horizon 2020 research and innovation programme (Grant agreement No. 101020057). R.R. acknowledges support from Grant RYC2021-030837-I, funded by MCIN/AEI/ 10.13039/501100011033 and by “European Union NextGeneration EU/PRTR”. This research was partially supported by the Spanish MINECO grant, PID2023-148661NB-I00. M.A.H. acknowledges a Warwick Astrophysics prize post-doctoral fellowship made possible thanks to a generous philanthropic donation.

This paper includes data collected with the TESS mission, obtained from the MAST data archive at the Space Telescope Science Institute (STScI). Funding for U.S. Institutions for the TESS mission is provided by the NASA Explorer Program. Funding for the TESS mission is provided by the NASA’s Science Mission Directorate. 

\software{TGLC \citep{BrandtTGLC}, Lightkurve \citep{2018ascl.soft12013L}, jaxoplanet \citep{jaxoplanet}, starry \citep{starry}, scipy \citep{2020SciPy-NMeth}, astropy \citep{astropy:2022}, numpy \citep{harris2020array}, matplotlib \citep{Hunter:2007}}

\bibliography{gd492}{}
\bibliographystyle{aasjournal}

\end{document}